\documentclass[a4paper,fleqn,usenatbib]{mnras}

\usepackage{newtxtext,newtxmath}

\usepackage[T1]{fontenc}
\usepackage{ae,aecompl}

\usepackage{graphicx}	
\usepackage{amsmath}	
\usepackage{amssymb}	
\usepackage{epstopdf}
\usepackage{lineno}
\usepackage{xspace}
\usepackage{lineno}
\usepackage{multirow} 
\usepackage{longtable}
\usepackage{supertabular,booktabs}
\title[QPO in Mrk 501] {Blazar Mrk 501 shows rhythmic oscillations in its $\gamma$-ray emission}

\author[G. Bhatta]{
Gopal Bhatta,$^{1}$\thanks{E-mail: gopalbhatta716@gmail.com}\\
$^{1}$Astronomical Observatory, Jagiellonian University, ul. Orla 171, 30-244 Krak\'ow, Poland
}

\date{Accepted XXX. Received YYY; in original form ZZZ}

\pubyear{2015}

\begin{document}
\label{firstpage}
\pagerange{\pageref{firstpage}--\pageref{lastpage}}
\maketitle

\begin{abstract}
 Quasi-periodic oscillations (QPO) originating from innermost regions
of blazars can provide unique perspective of some of the burning issues in blazar studies including disk-jet connection, launch of relativistic jets from the central engine, and other extreme conditions near the fast rotating supermassive black holes. However, a number of hurdles associated with searching QPOs in blazars e.g., red-noise dominance, modest significance of the detection and periodic modulation lasting for only a couple of cycles, make it difficult to estimate the true significance of the
detection. In this work, we report a $\sim$ 330-day QPO in the Fermi/LAT observations of
the blazar Mrk 501 spanning nearly a decade. To establish consistency of the result, we adopted multiple approaches to the time series analysis and employed four widely
known methods. Among these, Lomb-Scargle periodogram and weighted wavelet z-transform represent frequency domain based methods whereas epoch folding and z-transformed discrete auto-correlation function are time-domain based analysis. Power spectrum response method was followed to properly account for the red-noise, largely inherent in blazar light curves. Both local and global significance of the signal were found to be above 99\% over possible spurious detection. In the context where not many $\gamma$-ray QPOs have been reported to last more than 5 cycles, this might be one of the few instances where we witness a sub-year timescale $\gamma$-ray QPO persisting nearly 7 cycles. A number of possible scenarios linked with binary supermassive black hole, relativistic jets, and accretion disks can be invoked to explain the transient QPO. 
\end{abstract}




\section{Introduction}

Blazars, the most powerful sources in the Universe, are a small sub-class of active galactic nuclei (AGN). They are believed to harbor monstrously giant black holes, ($\sim 10^8-10^{10}M_\odot$), squeezed within a small volume comparable to that of the Solar System \citep[see][for recent black hole imaging of M87 galaxy]{2019ApJ...875L...6E}. The surrounding accretion disk constantly feeds the black hole with tremendous amount of matter which makes the nuclei extremely bright and consequently outshine the whole galaxy. The central engine spews out matter nearly at the speed of the light in the form of relativistic jets directed towards the Earth. Blazars consists two kinds of sources: flat-spectrum radio quasars (FSRQ), the ones which show emission lines and have the synchrotron peak in the lower electromagnetic frequency (typically $< 10^{14}$ Hz); and  BL Lacertae (BL Lac) objects, the others which  show weak or no emission lines and have synchrotron peak in the higher frequency. Although less powerful than FSRQs, BL Lacs are considered an extreme class of sources with highest synchrotron and inverse-Compton energies. More recently, blazar TXS 0506+056 was associated with the first non-stellar neutrino emission detected by the IceCube experiment \citep{2018Sci...361.1378I,2018Sci...361..147I, Padovani2018}.

 Blazars being the most luminous sources are \emph{visible} across the  universe;  and therefore, they can be used to probe the large scale  geometry of the universe.  Moreover, blazars could form an ideal sample of sources to study some of the challenging issues related to AGN e.g., interaction between accretion disk and jets, and the conditions leading to the launch of jets around supermassive black holes.  Although blazar cores mostly at large distances are not resolved by any current instruments, the presence of the  relativistic jets in blazars ``highly exaggerates'' the variability amplitudes and the timescales through relativistic beaming effects \citep[see][]{up95} such that information from the central engine can be carried along to us. For the reason,  variability studies can  be one of the most powerful methods -- if not the only method -- that guide us to the innermost regions of AGN, otherwise completely hidden from our view.

\begin{figure*}
\begin{center}
\includegraphics[width=0.98\textwidth,angle=0]{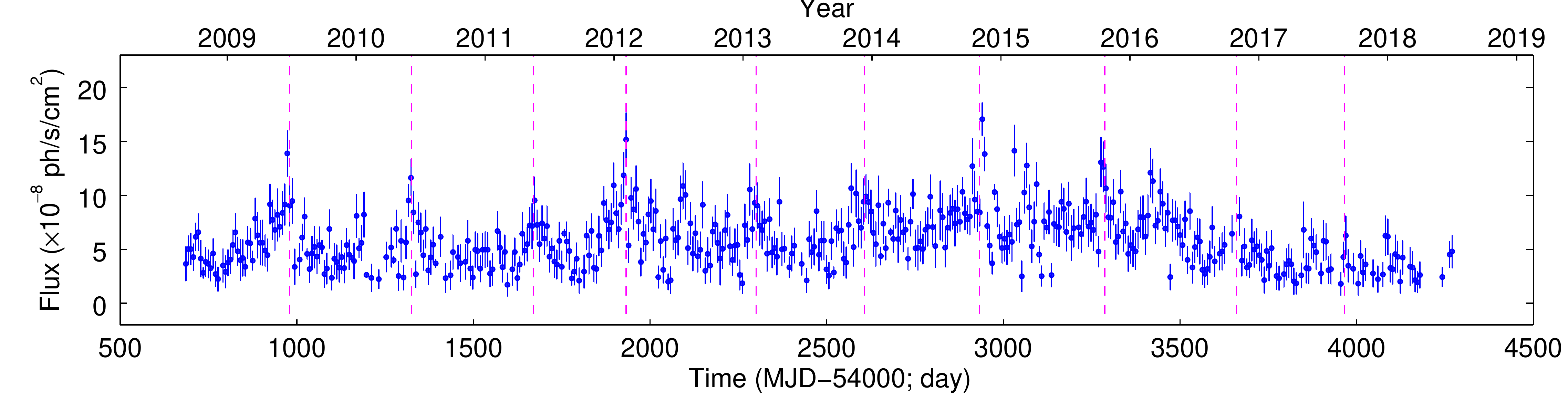}
\caption{The longterm Fermi/LAT  (0.1--300 GeV) weekly binned observations of the blazar Mrk 501. The vertical magenta lines approximately mark the peaks of the periodic oscillations (see Section \ref{sec:analysis}).} \label{Fig:1}
\end{center}
\end{figure*}

  Although, the flux variability shown by AGN, in general, appears to be aperiodic, the statistical nature of the observed variability can be broadly characterized as \emph{red-noise}, meaning source flux changes by larger amplitude over longer period of time.  However, over the past decade presence of quasi-periodic oscillations (QPO) in the multi-frequency blazar light curves has been recorded. The reported periodic timescales range from a few hours to several years \citep[see][and references therein]{Bhatta2017}. In particular, detection of QPOs in the $\gamma$-ray light curves of  a small number of blazars are intriguing. 
Of them, the famous case of 2.18-year periodicity in the TeV blazar PG 1553+113 was first reported in \citealt{Ackermann2015} and more recently in \citet{Tavani2018}. In addition, several works have claimed presence of $\gamma$-ray QPOs in other sources as well: \citet{Sandrinelli14} claimed $\sim1.73$-year QPO in the $\gamma$-ray light curve of the blazar PKS 2155-304 which later was confirmed by \citet{Zhang2017c}.
Similarly, \citet{Zhang2017a} reported 3.35-year periodicity in the blazar  PKS 0426-380 in 8-year long observation period.
 \citet{Zhang2017b} claimed 2.1-year QPO in the 9-year long observations of the source PKS 0301-243. 
In 8-years long $\gamma$-ray observations, \citet{Sandrinelli2017} studied QPO in three blazars S5 0716+714, Mrk 421, BL Lac and concluded that the first two blazars do not show any periodicity while BL Lac exhibited  a $\sim$1.86-year QPO,  similar to the one found in optical observations the source. It is important to note here that in all of these cases the claimed periodic timescales were nearly two years and above (or low temporal frequency), and number of observed periodic cycles were less than 5.  More recently, \citet{Covino2018} studying several blazars in $\gamma$-ray reported no significant periodicities in the light curves.

Also \citet{Sandrinelli2016b}  studied a number of blazars in the $\gamma$-ray band, including PKS 2155-304, OJ 287, and 3C 279   PKS 1510-089  and claimed the presence of year-like QPOs with modest significance. Similarly, \citet{Prokhorov2017}  in their search for the $\gamma$-ray sources showing periodicities confirmed previously reported quasi-periodicities in the three blazars, PG 1553+113, PKS 2155-304 and BL Lacertae, and  further observed evidence for quasi-periodic behavior of 4 blazars: S5 0716+716 and three of the high redshift blazars -- 4C +01.28,  PKS 0805-07 and PKS 2052-47.
 
   Mrk 501 (R.A.=$\rm 16^{h}53^{m}52.2^{s}$, Dec.= $+39\degr45\arcmin37\arcsec$) is one of the nearest blazars  (z = 0.034; Ulrich et al. 1975)  that shines bright in the X-ray.  Together with Mrk 421, it is one of the earliest extragalactic sources detected in the TeV band  \citep{Quinn1996}. Based on the location of the synchrotron peak, it is classified as high synchrotron peaked BL Lacs (HSP; \citealt{Abdo2010}). The source has been  extensively studied in a broad range of the electromagnetic spectrum by using several instruments including MAGIC, VERITAS, Whipple 10m Telescope, Fermi/LAT, RXTE and Swift, and radio and optical wavelengths \citep[][and references therein]{Ahnen2017,Furniss2015}.  A 23-day period was claimed to be found in the multi-frequency observations \citep[see][and references therein]{Rieger2000}, and later a harmonic with 72-day period was also reported by \citet{Rodig2009}.

 We organize the contents of the paper as following:  In Section \ref{sec:obs}, Fermi/LAT data acquisition and analysis are discussed. In Section \ref{sec:analysis}, we present the time series analysis of the blazar Mrk 501 using decade long $\gamma$-ray observations. Although we analyzed the light curve using four methods,  to avoid lengthy discussion on methods and put emphasis on more popular methods, we discuss  Lomb-Scargle periodogram (LSP) and the weighted wavelet $z$-transform (WWZ) methods in the main paper while we move the epoch folding and   z-transformed discrete auto-correlation function analysis to Appendix. 
 In the section, we report a $\sim$ 330-day quasi-periodic modulations in the $\gamma$-ray light curve. We also note that the periodic signal gradually weakened near the end of the observation period.  Finally, in Section  \ref{discussion} we discuss some of the possible scenarios that can explain the observed periodicity.

\section{Observations and Data Reduction }
\label{sec:obs}
  The Large Area Telescope (LAT) of the Fermi Gamma-ray Space Telescope  operates between 20 MeV to 300 GeV energy range and it continuously scans the sky \citep{Atwood2009}. The telescope has been playing a crucial role to the study of the Universe  since its launch in the year 2008.  For the periodicity  analysis, $\gamma$-ray observations of the source 3FGL J1653.9+3945 spanning $\sim 10$ years (54682 -- 58270 MJD) were used. The observations in the energy range 10 MeV -- 300 GeV  were  processed following the standard procedures of the unbinned likelihood  analysis \footnote{\url{https://fermi.gsfc.nasa.gov/ssc/data/analysis/scitools/}}. The data were analyzed using Science Tools version v11r5p3\footnote{\url{https://fermi.gsfc.nasa.gov/ssc/data/analysis/software/v11r5p3.html}} and the LAT  instrument response function  P8R2\_SOURCE\_V6. In particular, only the events within a circular region of interest (ROI) of $10^{\circ}$ radius centered around the source were considered; and to avoid the contamination of $\gamma$-rays from the Earth limb the zenith angle was restricted to less than 90$^{\circ}$.  Likelihood analysis \citep{Mattox1996} was carried out using all the point sources in the ROI along with the diffuse emission background models  - Galactic (gll\_ iem\_v06.fit) and extra-galactic (iso P8R2\_SOURCE\_V6\_v06.txt)\footnote{\url{https://fermi.gsfc.nasa.gov/ssc/data/access/lat/BackgroundModels.html}}. In likelihood analysis  the test statics (TS) for the null hypothesis  asymptotically follows  $\chi^2_n$  distribution, where n  is the number of parameters characterizing the additional source at a location. In the course of Fermi/LAT analysis, each fit at a given position of the sky involves two degrees of freedom (i.e. flux and spectral index). Therefore, the TS  nearly following $\chi^2_2$ distribution,  significance of  an event can be given by $\sim \sqrt{TS} \sigma$ (\citealt{Abdo2010}; details on the Fermi/LAT data processing are also discussed in \citealt{Bhatta2017}). To construct the light curve, a graph that shows photon flux variation as a function of time,  we binned the observations in a  7-day  bin. For a robust time series analysis, the study includes only the observations with TS value $>$ 10 (or above $3\sigma$ significance), of which more than 80\%  observations are above 5$\sigma$ significance level.

\section{Analysis and Results }
\label{sec:analysis}

 The Fermi/LAT  light curve of the blazar Mrk 501 is presented in Figure \ref{Fig:1}  which shows a clear trend of quasi-periodic oscillations in the $\gamma$-ray emission. The observed mean long-term variability was measured as  fractional variability \citep{vau03,Bhatta2018a} of  $33\pm4$\%. To search for  the possible QPOs in the Fermi/LAT light curve, Lomb-Scargle periodogram (LSP;  \citealt{Lomb76,Scargle82}) 
 was performed. The method  approaches periodicity analysis from multiple perspectives and thereby contains some elements of several other methods e.g. \emph{Fourier analysis}, \emph{least-squares method},  \emph{Bayesian probability theory},  and  \emph{bin-based phase-folding} techniques \citep[see][]{VanderPlas2018}.

\begin{figure}
\begin{center}
\includegraphics[width=\columnwidth]{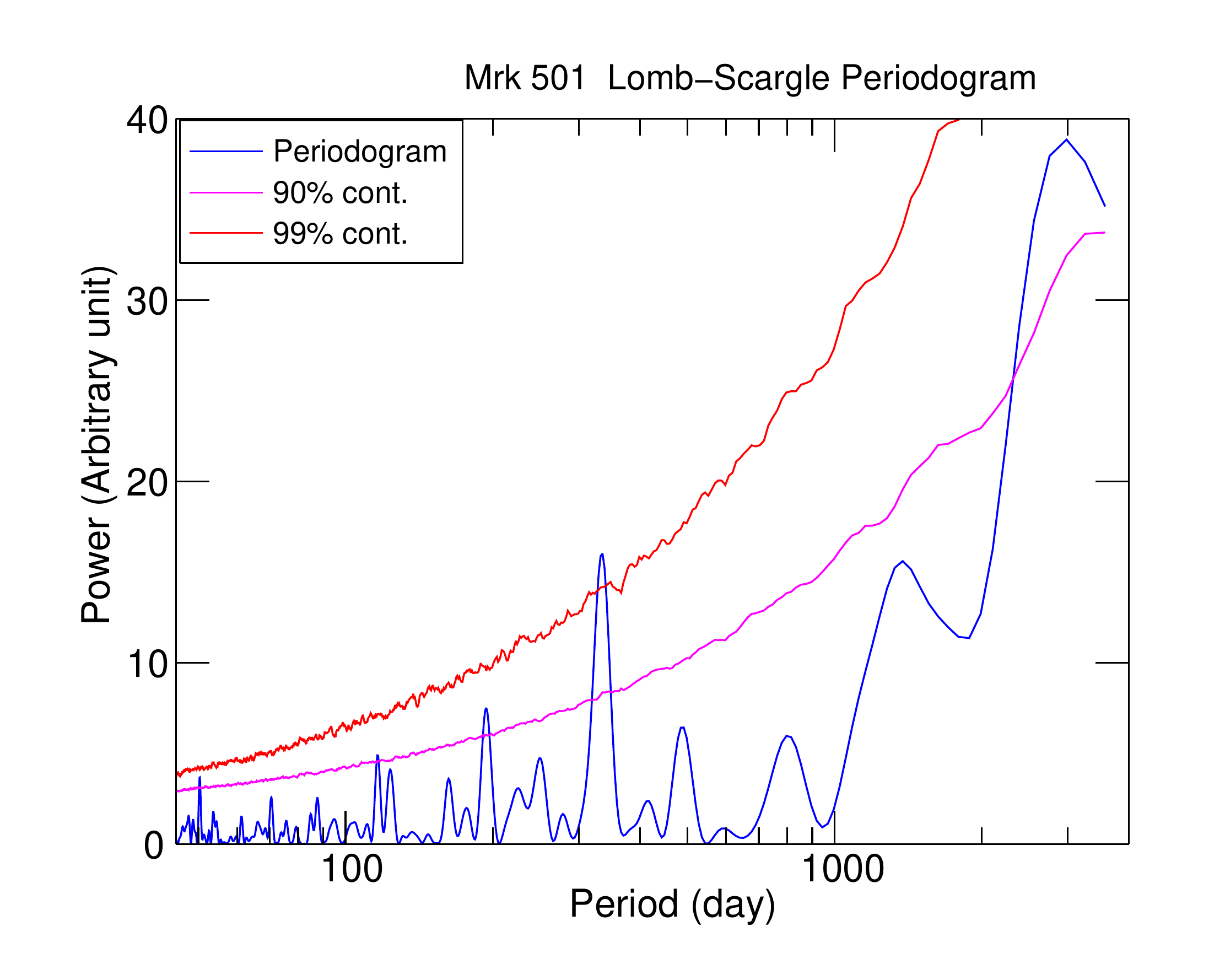}\\
\caption{The Lomb-Scargle periodogram of the Fermi/LAT light curve of the source Mrk 501 is represented by the black curve. The magenta  and the red curve show the local 90\%  and  99\% significance contours, respectively, from the MC simulations. \label{Fig2}}
\end{center}
\end{figure}

The  Lomb-Scargle (LS) periodogram of the source $\gamma$-ray light curve was computed for the temporal frequencies between the minimum $f_{min} = 1/3584$ d, and  maximum $ f_{max}$=1/14 d.  The number of frequencies in-between were chosen so as to gain a balance between the spectral resolution and computational convenience \citep[see][]{Bhatta2018d}. The periodogram is presented in Figure~\ref{Fig2} which shows a distinct peak  around the timescale of $332\pm17$ days suggesting  presence of a periodic signal at the timescale. The half-width at the half-maximum (HWHM) of the peak was taken as a measure for the uncertainty in the observed period. To better \emph{see} the periodic oscillations in the light curve in Figure \ref{Fig:1}, vertical lines in magenta color are drawn approximately at the peak of the oscillations corresponding to the detected 332-day period.

However, it is important to point out that, in general, the statistical properties of  blazar light curves can be characterized as red-noise processes such that the light curves dominated by such noise can occasionally mimic a  periodic behavior for  a few-cycles,  especially in the low-frequency (i.e. longer timescale) regime \citep[see][for the discussion]{Vaughan2016,Press78}. Therefore while estimating the significance of any  periodic feature observed in a periodogram, the red-noise behavior should be properly accounted along with the other effects due to the discrete sampling, finite observation length and uneven sampling of the light curve.  
  
 \begin{figure}
 \includegraphics[width=\columnwidth]{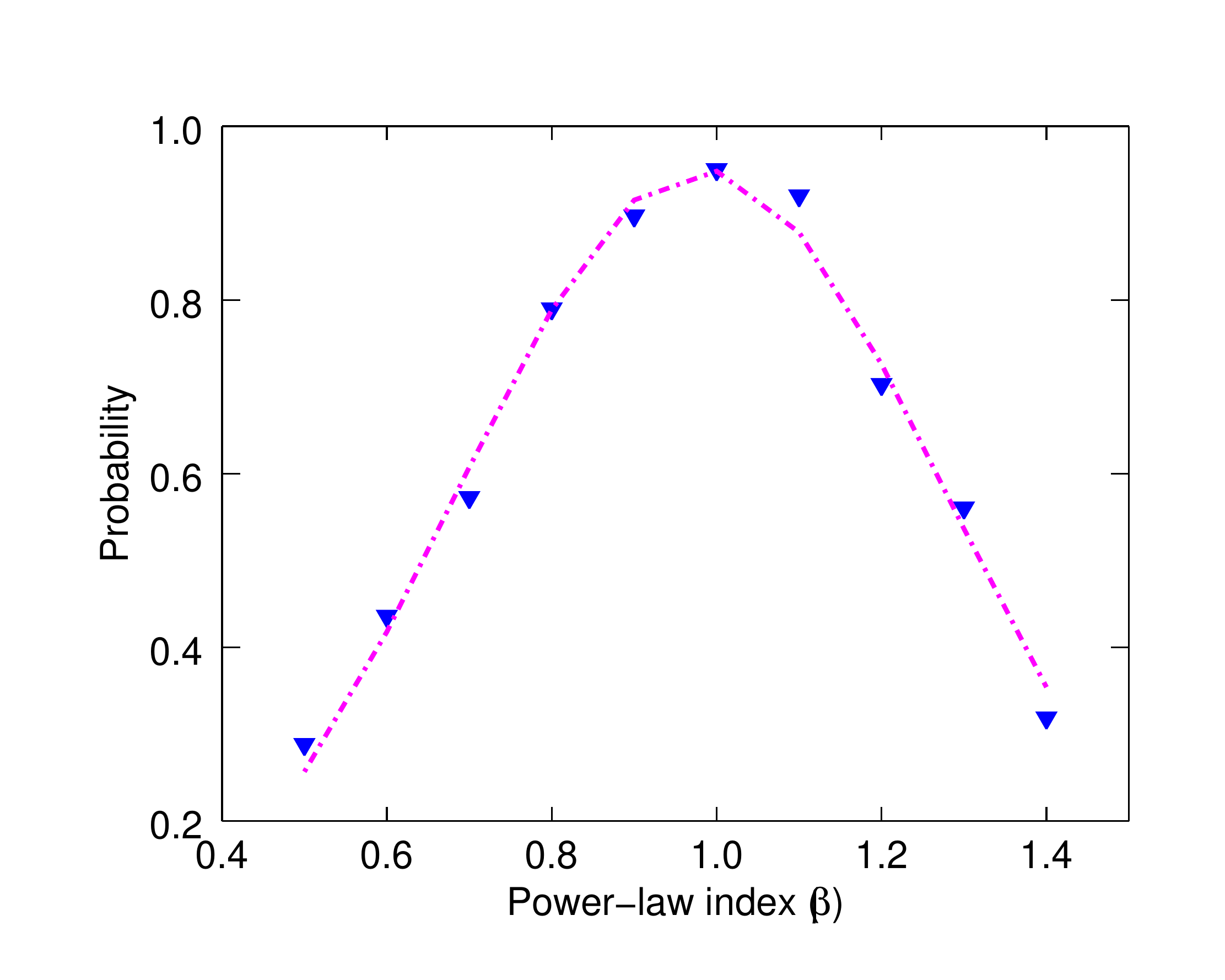}
 \includegraphics[width=\columnwidth]{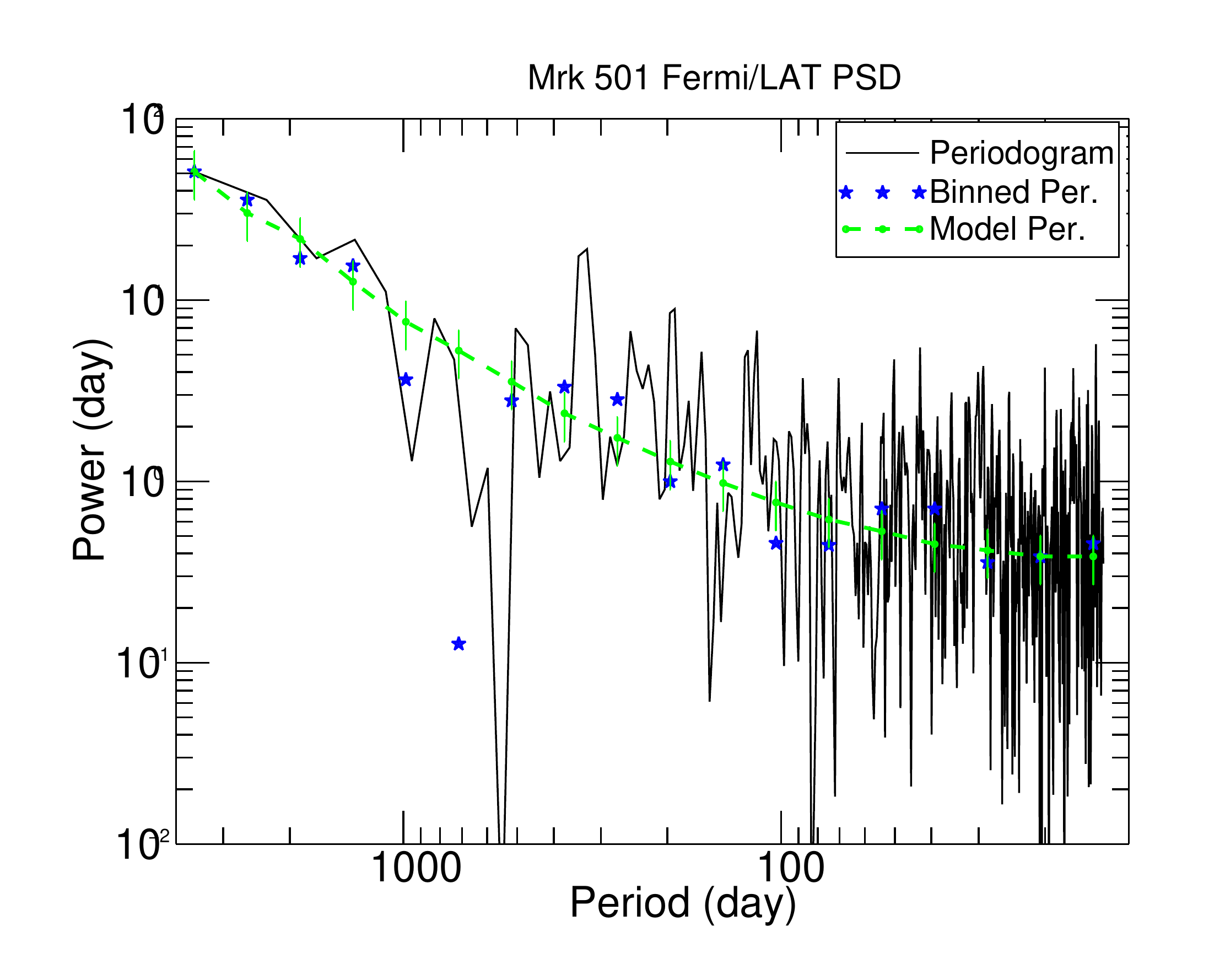}
\caption{ Top: Distribution of the probabilities over the power-law indices of the model PSD representing the periodogram from the method PSRESP (see Appendix 2). Bottom: Discrete Fourier periodogram  of the Fermi/LAT light curve of the blazar Mrk 501 shown in the black color. Blue symbols represent the binned periodogram and the best-fit model PSD is shown by the green curve. The errors are estimated using Monte Carlo simulation. \label{Fig3}}
\end{figure}  

 \begin{figure*}
\begin{center}
\includegraphics[width=\textwidth]{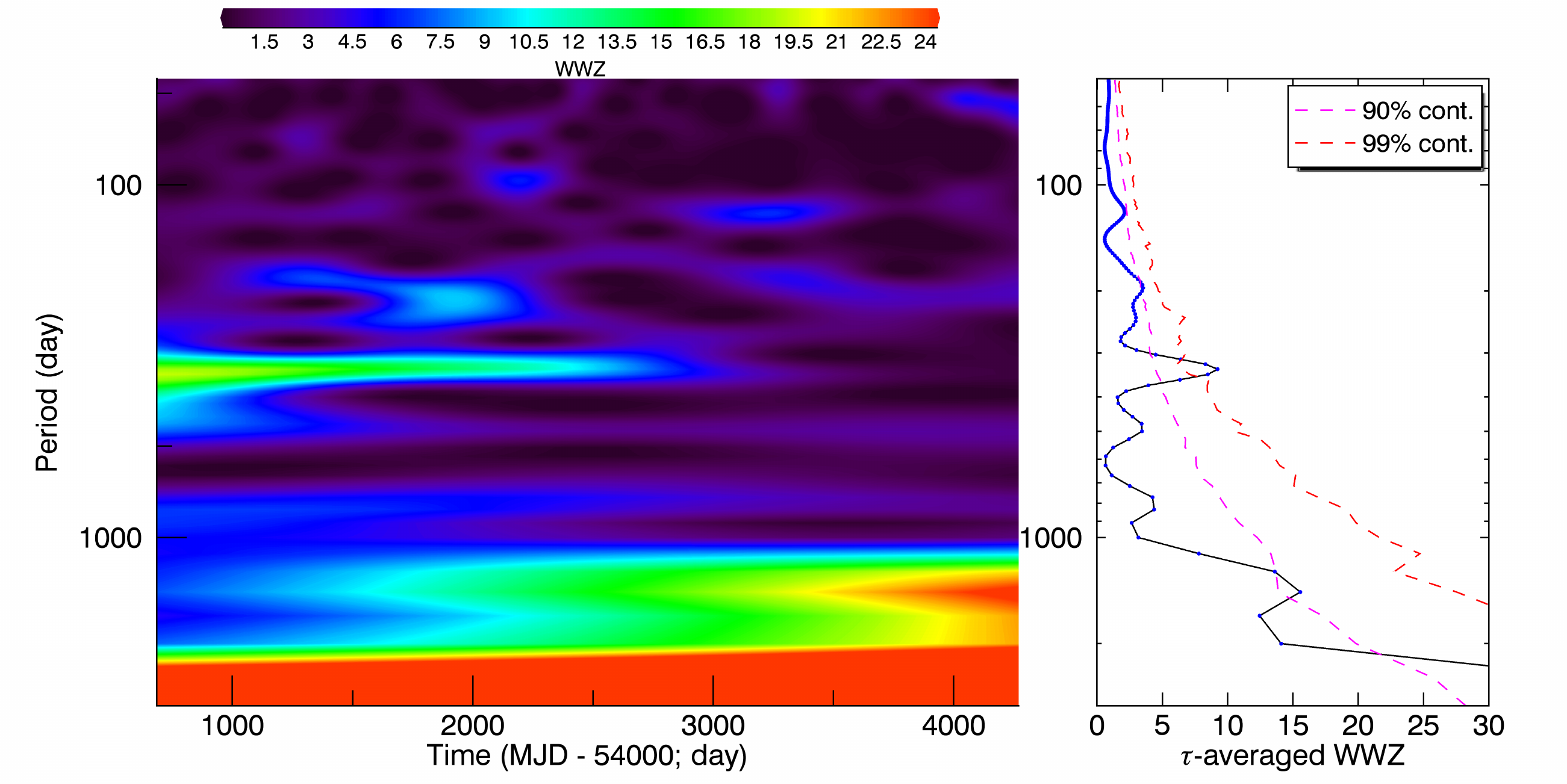}
\caption{Weighted wavelet z-transform  of  $\gamma$-ray light curve for the blazar Mrk 501. The left panel shows the distribution of color-scaled WWZ power in the time-period plane, and the right panel shows the time averaged power (blue curve) as a function of period. The magenta and the red curves show the 90\%  and 99\% local significance contours, respectively,  from the MC simulations. \label{Fig4}}
\end{center}
\end{figure*}

 To quantify the significance of the observed feature against  spurious detection,  the power spectrum response method  \citep[PSRESP;][]{Uttley02} was followed,  and see also \citep[][]{Edelson2014,bhatta16b,bhatta16c,Bhatta2017,Bhatta2018b}.  We first modeled the source periodogram  with a power-law PSD of the form $P(\nu) \propto \nu ^{-\beta}+C$; where ${\nu}$, ${\beta}$, and $C$ represent temporal frequency, spectral index and Poisson noise level, respectively. To estimate the best fit power-law index, 10000 light curves were simulated by the method described in  \citet[][]{TK95} for each of the index values from 0.5 to 1.4 with an interval of 0.1. Then the probability that the PSD shape represents the source periodogram was calculated as described in Appendix 2. The resulting probability distribution over the index values is presented in the top panel of Figure \ref{Fig3}, and the best-fit index with the highest probability 0.94 is estimated to be $0.99\pm0.01$ from the Gaussian fit of the probability distribution. For $N$ number of observations during the observation period $T$ when mean flux is $\mu$, the Poisson noise level can be given by
\begin{equation} 
 C=\frac{2T}{N^2\mu^2}\bar{err^2},
\end{equation} 
where $\bar{err^2}$ stand for mean square of the flux uncertainties. With $C=3.0\times10^{-3}$ d and  $\beta=0.99$, the best-fit PSD model (green) is shown overlaid on the discrete Fourier periodogram of the source light curve in the bottom panel of Figure \ref{Fig3}. 

Once the best-fit PSD shape was configured,  10000 light curve were simulated using the shape and their LS periodogram was computed. Consequently, using the spectral distribution of the simulated light curves  local 90\% and 99\% significance contours were evaluated \citep[for further details see][]{bhatta16b,bhatta16c,Bhatta2017,Bhatta2018b}.  In Figure \ref{Fig2}, the 90\% and 99\% significance contours are shown by the magenta and the red curves, respectively.  As we can see, the $\sim$332-d QPO with $\sim$99.4\% local significance has high a probability of being a real periodic signal. Note that there are also two other LSP peaks, at 195 d and 76 d, respectively, that are slightly above 90\% contour, and one feature at 116 d period that  just touches the 90\% contour.  However, as these peaks stand below well 99\% significance, they are less likely to represent real periodicities. Therefore, we will not further discuss about these peaks in this work.

For the main peak, we also estimated global significance which is devised to address mainly two issues:  1) We do not have \textsl{a priori} knowledge of the period in which the significant peak might occur, and 2)  the local significance does not consider the number of independent frequencies used in the computation. Therefore, the global significance is defined as a fraction of periodograms at all timescales (alternatively, frequencies) considered that are below the observed feature at the particular period (for similar definition of global significance in the context of cross-correlation between two light curves, refer to \citealt{Bell2011}).  Again, utilizing the distribution of simulated LS periodogram using best-fit PSD model, we evaluated the global significance in reference to the LSP peak at the period $\sim$332 day to be 99.1\%.  

 To further investigate the transient nature of the above QPO in the $\gamma$-ray light curve of the blazar Mrk 501, weighted wavelet z-transform method (WWZ; \citealt{Foster96}) \citep[see also][]{Gupta2018,Bhatta2017} was used to estimate the WWZ power of the Fermi/LAT observations. Figure~\ref{Fig4} presents the WWZ powers of the source light curve  color-scaled in the time-period plane. We find that  significant WWZ power are centered near 332-day period, which further confirms the detection of the QPO by the LSP method. The Figure also shows that the QPO starts to decay from the time $\sim3000$ days from 54000 MJD (equivalently, $\sim2300$ days after the start of the observation) and gradually disappears towards the end of the observation period. Such a behavior is also visible in the light curve in Figure \ref{Fig:1} which clearly shows the amplitude of modulations grow weaker after the year 2016.  The right panel of Figure~\ref{Fig4} shows the time-averaged WWZ power at a given period. Once again, in the panel, we can see a distinct peak centered around the periods of $332\pm28$ days. As in the LSP, HWHMs about the central peaks provide a measure for the uncertainties in the observed period. 

 In order to evaluate the significance of the WWZ detected quasi-periodic feature at the timescale of $\sim$332 days, we took the approach that is similar to the one adopted in case of the LSP.  The best-fitting PSD model was used to simulate 10000 light curves which subsequently were re-sampled to match the sampling of the source light curve. The distribution of the simulated time averaged WWZ  powers were then utilized to evaluate  the  90\%  and the 99\% local significance  contours as shown in the right panel of  Figure \ref{Fig4} by  the magenta and the red curves, respectively \citep[see][for further details]{Bhatta2017}.  The local and global significance of the QPO were estimated to be 99.7\% and 99.2\%, receptively. We note that a slightly larger local significance, in comparison to the one in the LSP method, is consistent with the fact that wavelet analysis looks for periodic behavior localized in frequency and time. 
 
  To robustly establish consistency of the QPO significance irrespective of the types of methods used, in addition to LSP and WWZ, which are frequency domain based analysis, we also performed two time domain based analyses, namely, z-transformed discrete auto-correlation function (ZACF) and epoch-folding. The results of the analyses which confirm the presence of the $\gamma$-ray QPO are presented in the Appendix 1.

\section{Discussion and Conclusion }
\label{discussion}

We analyzed $\sim$10 year long Fermi/LAT observations of the famous blazar Mrk 501 using LSP, WWZ, epoch-folding and ZACF methods. All four methods consistently detected a transient quasi-periodic oscillations of the period centered around 332 days. The period in the source rest frame ($P$) is  estimated as $P=P_{obs}/(1+z)$ $=321$ days, where $P_{obs}$ and $z$ are the observed period and the red-shift, respectively. The observed transient feature might have arisen due to correlation of the noise e.g. colored (power-law) noise in the light curve. Therefore, we employed both frequency and time domain based analysis, and subsequently, estimated the local and global significance of the observed period taking account a number of artefacts that potentially could produce spurious peaks e.g. red-noise, discrete data, finite observation length, and uneven data sampling. We simulated a large number of light curves from the best-fit PSD model and subsequently assigned the sampling of the real observation.  The significance (both local and global) of the detection in LSP and WWZ methods was found to be above 99\%.  In other words, the probability that observed periodicity was generated due to correlation of noise was $<1\%$ and therefore, the feature at the timescale of 332 days is highly likely to be of the physical origin. A quick analysis of the source was also performed in the optical waveband using the publicly available data from the Catalina Real-time Transient Survey\footnote{\url{http://nesssi.cacr.caltech.edu/DataRelease/}}.  Some hints of the $\gamma$-ray detected periodicity was also found in the analysis. However,  the optical observations being sparse with large gaps in between, we could not be conclusive about the results.  A  multi-frequency time series analysis of the source is deferred to the future study.

It should be pointed out that most of the $\gamma$-ray QPOs mentioned in the Introduction are reported to have lasted only a few cycles (typically less 5 cycles). Against such a backdrop, what we have observed might be the first interesting event where we witnessed transient $\gamma$-ray periodic modulations in the  high energy emission of the blazar that  persist  up to 7 cycles before the oscillations gradually fade away. It  could be also possible that the flux modulations started long before the Fermi/LAT launch year (i.e. 2008).  In that case, the modulations might have spanned more than 7 cycles. In any case, the detection adds Mrk 501 to the list of a few reported cases of QPOs in blazar $\gamma$-ray light curves.

  Detection of such periodic signals is immensely useful because the signals, mostly originating from the innermost regions of the blazar cores, carry information about the inner structures including space-time geometry around fast spinning supermassive black holes, disk-jet connection, and magnetic field configuration of the accretion disk. QPOs in blazars can be linked to a host of possible events that are chiefly based on three scenarios: supermassive binary black holes (SMBBH), accretion disk based instabilities, and events related to geometric configuration of the jet. Below we discuss some of these in the context of the observed QPO in Mrk 501.
  
  \begin{itemize}
  \item Supermassive binary black holes:  Periodic $\gamma$-ray flux modulations are most likely to originate from the relativistic jets, but we may still need a periodic perturbation close to the central engine which would propagate along the jet.  It is found that SMBBH systems provide the most natural model for such recurring  perturbations \citep[e.g.,][]{Lehto96,Valtonen08} that give rise to periodic flux modulations.  In Mrk 501, a 23-day periodic behavior previously reported was based on this model \citep[see][]{Rieger2000}. Interestingly, in both the cases the maxima on average are nearly 8 times the minima of the modulations. The SMBBH orbit can fairly be assumed to be circular as dynamical friction gradually tends to erase any initial eccentricity over the course of merging. The radius of the Keplerian orbit can be calculated using the relation,
  \begin{equation}
r=9.5\times 10^{-5}\left ( \frac{M}{10^9M_\odot } \right )^{1/3}\tau _k^{2/3} \rm \ pc
 \end{equation}

In the case of Mrk 501 which has a black hole of mass $7\times10^8M_\odot$ \citep{Ghisellini2010}, the total mass of the system can safely be assumed to be $1\times10^9 M_\odot $. Then the black holes can be estimated to have a separation of $\sim 5$ milli-parsecs. However, such a close SMBBH systems has not been detected yet. 

 If the jet axis is not aligned to the total angular momentum vector of the binary system, it will induce precession in the trajectory leading to  accretion disk precession under gravitational torque \citep[e.g][]{Katz1997} or the
 jet precession  \citep[e.g.][]{Begelman1980,Caproni2017,Sobacchi2017},  and also precession in periodic impact as in  OJ 287  \citep{Sillanpaa88}. In the case of significantly elliptical binary orbit, the secondary BH periodically perturbs the primary jet at periastron to induce magnetohydrodynamic (MHD) instabilities. These, in turn, can lead to  particle acceleration via magnetic reconnection events. Consequently, the particles emit via synchrotron and inverse Compton emission \citep[see][for details]{Tavani2018}. In the case of Doppler-boosted flux modulations, the true period in the source rest frame takes the form $P_{obs}=(1+z)(1-\beta cos\theta)\tau_k$, so that for a typical bulk Lorentz factor of 10, the corresponding Keplerian orbit at the radius of 0.1 pc has a period of $\sim 90$ years. \citep[For illustration, refer to the left panel of Figure 4 in][]{Bhatta2018d}
  
  \item Accretion disk based events:  QPOs might arise simply due to hot-spots on the accretion disk revolving around the black holes. In such cases, taking the black hole mass $7\times10^8M_\odot$ for the source, the radius of the corresponding hot-spot turns around to be about 90 gravitational radii $r_g$ or about 4 milli-parsecs, where $r_g=GM/c^2$. Similarly, tilted inner disks around spinning black holes undergo Lense-Thirring precession resulting in periodic flux modulations\citep[e.g.][]{Stella98}.  The Lense-Thirring effect around a spinning black hole with Kerr spin parameter $a$ causes tilted disk orbits to precess about its angular momentum vector with a period conveniently expressed as    
    \begin{equation}
\tau_{LT}= \frac{1}{2a} \left ( \frac{r}{r_g} \right )^{3} \tau_{k-r_g},
\end{equation}
  \noindent  where  $\tau_{k-r_g}$ is the Keplerian period at the black hole's gravitational radius, $r_g = GM/c^2$. In such scenario, precession around fast spinning black hole ($a=0.9$), with the given mass of the  source, the observed period in the source rest frame can be used to estimate the inner radius of the orbit to be $\simeq 13\ r_g$. When the disk is strongly coupled to the jet, the precession of the disk can make the jets precess. Recently \citet[][]{Liska2018} performed 3D general relativistic MHD simulations  based on the similar scenario; the results showed that the jets exhibit precession with a period of $\sim 1000 t_g$, where $t_g=r_g/c$, which is of the order of the period we observed in the source. 
  
  Similarly, in the case of globally perturbed thick accretion disks \citep[e.g.][]{Rezzolla2003,Zanotti2003}, the disk undergoes p-mode oscillations such that the fundamental frequency of the oscillation can be approximated as,  
   \begin{equation}
  f_{0}\approx 100\left ( \frac{r}{r_g} \right )^{-3/2}\left ( \frac{M}{10^8M\odot } \right )^{-1} \ \rm day^{-1}
 \end{equation}
 \citep[see][]{Liu2006,An2013}.  If the observed 321 rest-frame period is taken to be the period corresponding to such fundamental frequency, the inner radii of the perturbed disks turns out to be $\sim 276\ r_g$. Interestingly, this represents the order of the inner radius of the truncated disk in some of the radio loud bread-line region galaxies showing disk-jet coupling \citep[see e.g.,][]{Bhatta2018b}. In the sources showing strong disk-jet connection, such perturbations could propagate along the jet resulting in high energy QPOs.

  \item Jet Geometry: The transient 332\--day period might as well be linked to geometrical configuration of the jet, such that instabilities propagating with relativistic speeds along the magnetized jets e.g., following a helical path \citep[as in][]{Camenzind92,Mohan15}, could provide a possible explanation for the QPOs in blazars. In such a scenario, for the emission regions moving with a typical bulk Lorentz factor $\Gamma \sim 10$ along the trajectory viewed about an angle, $\theta \sim 1/\Gamma$, the time period in the source rest frame, corresponding to 332 days observed period, can be estimated using  $P\simeq \Gamma ^2P_{obs}/(1+z)$. The calculated timescale is in the order of one 100 years, and it provides us an upper limit of a few tens of parsecs to the spatial scale for the periodic structures. Alternatively, perturbations close to the central engine propagate along the jet inducing periodic swings in the jet viewing angle. Using Doppler boosting relation $F(\nu)=\delta^{(3+\alpha)}F'(\nu')$, for a given change in the angle $\Delta \theta$, the ratio of the observed  flux modulation, intrinsic flux remaining unchanged,  can be given as
 \begin{equation}
 \Delta logF=-\left ( 3+\alpha  \right )\delta \Gamma \beta sin\theta \Delta \theta ,
 \end{equation}
where negative sign indicates that for a positive change in the angle, representing the emission region moving away from the line of sight, the flux ratio decreases. 
 In such events, periodic changes in the viewing angle as small as 1$^{\circ}$ can also give rise to periodic flux modulations by an order of magnitude \citep[For illustration for flux doubling case, refer to the right panel of Figure 4 in][]{Bhatta2018d}.
    \end{itemize} 
 
We also note the interesting feature in the light curve (see Figure \ref{Fig:1}) that the QPO fades around the same time when the  $\gamma$-ray  emission diminishes in intensity. This might indicate a close link between QPOs and jet production mechanisms in blazars. Such possible link can still be explained in the scenarios discussed above. For example, the gravitational perturbation in a SMBBH system can feed to the instabilities \emph{in situ} that contribute to the $\gamma$-ray  production and consequently, result in transient QPOs; the disk hot-spots influencing jet emission can frequently form and disrupt on some characteristic timescale. Similarly, the jet can turn away reducing the  $\gamma$-ray  emission and alter the periodicity at the same time via Doppler boosting. 
  Finally,  although all of these scenarios can lead to QPOs, without further multi-frequency time series analysis and collective discussion on the topic, the exact cause of the observed transient $\gamma$-ray rhythmic flux modulations would still remain mysterious.

\section*{Acknowledgment}
 I acknowledge the financial support by the Polish National Science Centre through the grant UMO-2017/26/D/ST9/01178.  I thank Dr. Niraj Dhital for the important discussion and useful comments. I am grateful to the anonymous referee for his/her comments that significantly improved the quality of the work.

\clearpage

\onecolumn
\section{Appendix 1: Epoch-folding and z-transformed discrete auto-correlation function \label{sec:app}}
We also explored the  periodicity in the source by using two more methods: epoch folding and z-transformed discrete auto-correlation function (ZDCF). The methods, unlike previously presented frequency based analysis, i.e. LSP and WWZ, are time-domain based analysis. Details on the usage of epoch-folding can be found in \cite{Bhatta2018d}, and ZDCF method is discussed in \citep{Bhatta2018c, Bhatta2018a}.

   \begin{figure}
\includegraphics[width=0.7\textwidth]{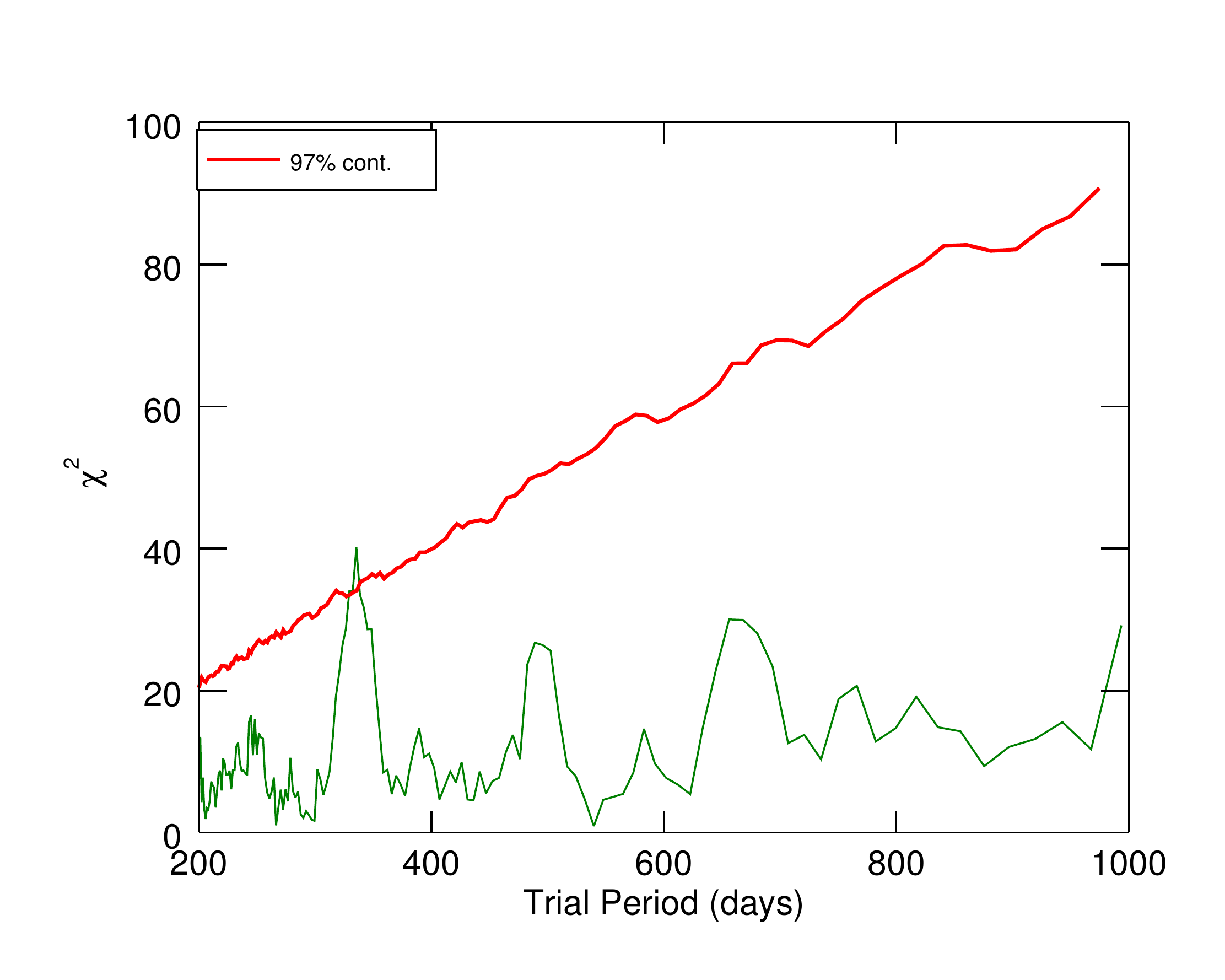}
\caption{Epoch folding of the long-term $\gamma$-ray light curve of the blazar Mrk 501. The figure shows $\chi^2$ values for the trial periods between 200 and 1000 days. The red curve represents the 97\% significance contour from the MC simulations using best-fit PSD model (see Section 3). The significance of the $\chi^2$ peak at  $\sim 330$ days trial period is estimated to be 98.7\%  \label{Fig5}}
\end{figure}

\section{Appendix 2: Power spectrum response method (PSRESP) \label{sec:app}}
The PSRESP method is aimed at estimating the power spectral density (PSD) shape that best represents periodogram of an unevenly spaced light curve. The method, based on intense simulations, properly takes account of a number of artefacts, usually unavoidable in frequency domain analysis, such as \emph{red-noise leak}, \emph{aliasing} and PSD distortion  due to unevenly spaced sampling. In addition, it provides a measure for goodness of fit and, thereby, estimates uncertainties in the model parameters. The method is fully described in \citealt{Uttley02} (see also \citealt{bhatta16b} and Appendix in \citealt{Chatterjee08}); here, for completeness, we briefly summarize the method as following:
\begin{itemize} 
\item[i)] For a given time series $f(t_{j})$ sampled at times $t_{j}$ with $j = 1, 2,.., N$ during an observational period $T$,  the periodogram at a given  frequency $\nu$ is estimated using the expression
\begin{equation}
P\!\left(\nu \right)= \left | \sum_{j=1}^{N}f\!\left( t_{i} \right ) \, e^{-i2 \pi \nu t_{j}} \right |^{2} \, .
\end{equation}
 Often, to express the variability power in terms of fractional variation, a normalization of the kind $N^2 \mu^2/2T$ is applied, where  $\mu$ stands for the mean flux during the entire observational period. The periodogram  is computed at regular frequency grid starting from the minimum frequency $\nu_{min}$=1/T to maximum (Nyquist) frequency $\nu_{max}=N/2 T$.

\item[ii)] To estimate power-law index that best represents a given periodogram,  a large number of light curves (typically 1000) are simulated based on a power-law model of the form $P(f) =  \nu^{-\beta }+C$, where $\beta$ and C stand for power-law index and the Poisson noise. Thus simulated light curves, subsequently, are re-sampled according to the source light curve before periodogram for each one of them is calculated.
\item[iii)] For each of the simulated light curve, a $\chi^{2}$-like quantity is constructed which is given as
\begin{equation}
\chi_{i}^{2}=\sum_{\nu _{\rm min}}^{\nu _{\rm max}}\frac{\left [ \overline{P_{\rm sim}\left ( \nu  \right )} -P_{i}\left ( \nu  \right )\right ]^{2}}{\Delta \overline{P_{\rm sim}\left ( \nu  \right )}^{2}},
\label{chi}
\end{equation}
\noindent where $ \overline{P_{\rm sim}\left ( \nu  \right )}$ and $\Delta \overline{P_{\rm sim}\left ( \nu  \right )}$ represent the mean periodogram and the standard deviation of the simulated periodograms; a similar quantity for the observed periodogram, $\chi_{\rm obs}^{2}$, is also evaluated by replacing $P_{i}$ with $P_{\rm obs}$.
\item[iv)] Step iii) is repeated for a number of indices of the power-law model.
\item[v)] The goodness of fit between the mean simulated periodogram and the observed periodogram is estimated by comparing $\chi^{2}_{\rm obs}$ with $\chi^{2}_{i}$s. In particular, to quantify the goodness of the fit for a given index, the ratio of the number of $\chi^{2}_{i}$s greater than $\chi^{2}_{obs}$ to the total number of $\chi^{2}_{i}$s in all simulations is defined as the probability that the model, with the given index, best represents the source periodogram.
\end{itemize} 

\begin{figure}
\centering
\begin{tabular}{c@{}c}
\hspace{-0.48cm}
\resizebox{0.52\textwidth}{!}{\includegraphics{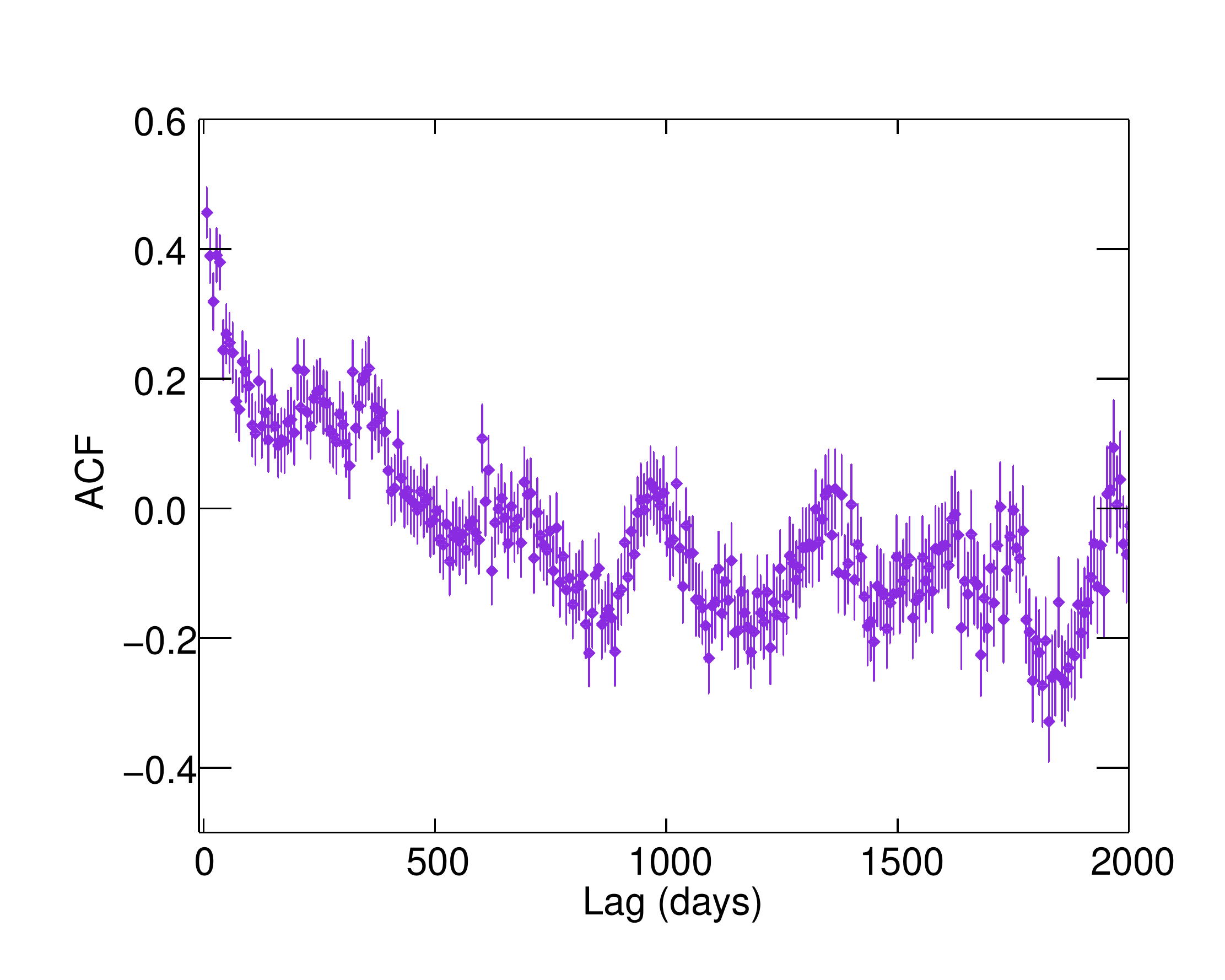}} &
\hspace{-0.48cm}
\resizebox{0.52\textwidth}{!}{\includegraphics{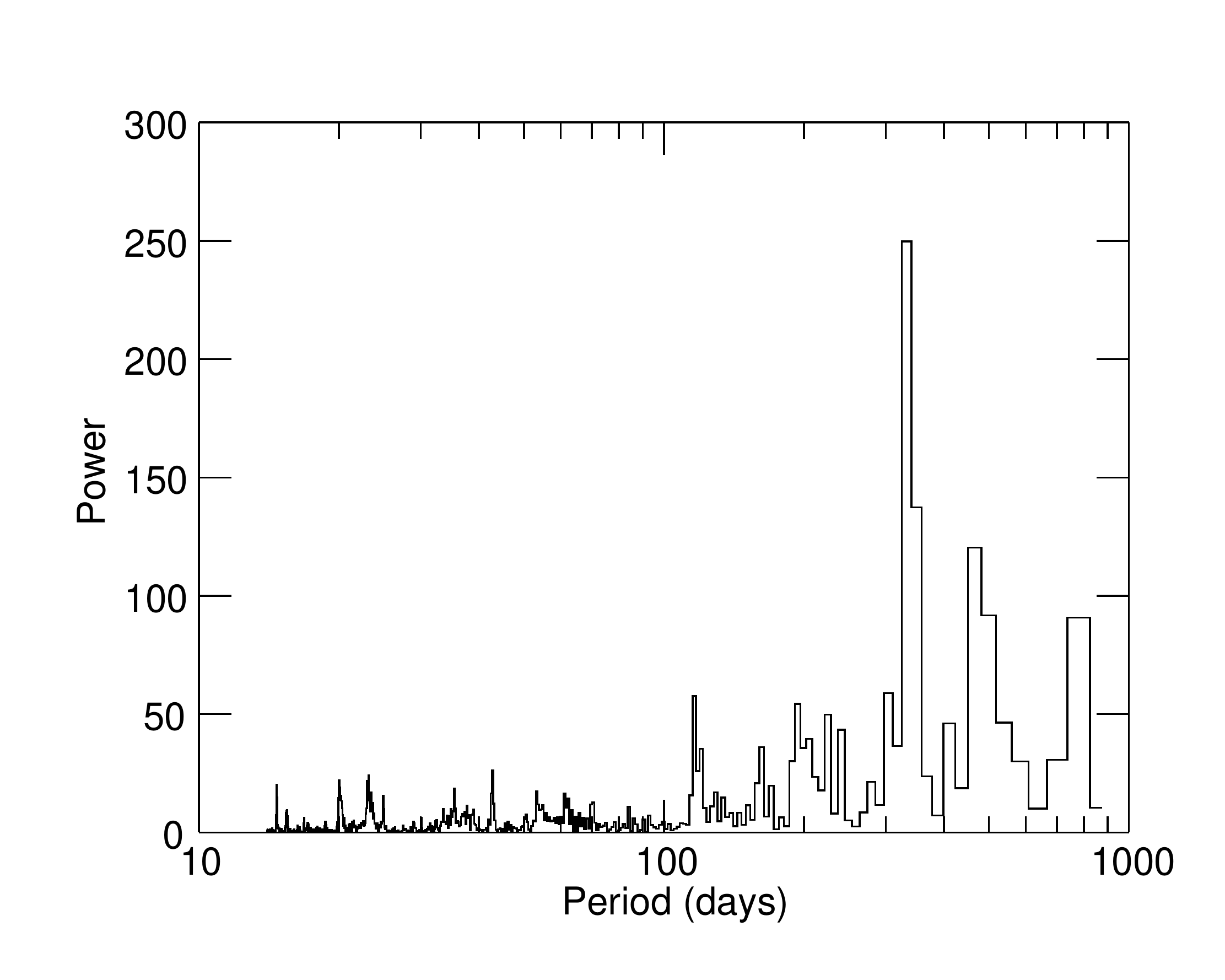}} \\
\end{tabular}
\caption{z-transformed  auto-correlation function (ACF) of the long-term $\gamma$-ray light curve of the blazar Mrk 501. Epoch folding of the $\gamma$-ray light curve of the blazar Mrk 501. Left panel: ACF displays maxima at the interval of $\sim 330$ days. Right panel: Discrete Fourier periodogram of the ACF curve showing the most prominent feature $\sim 330$ days period.}
\label{fold}
\end{figure}

\end{document}